\documentclass[12pt]{article}
\usepackage{amsfonts,amsmath,amssymb}
\usepackage{amssymb}
\usepackage{epsfig}

\begin{document}\def\p{\phi}\def\P{\Phi}\def\a{\alpha}\def\e{\epsilon}
\def\be{\begin{equation}}\def\ee{\end{equation}}\def\l{\label}
\def\0{\setcounter{equation}{0}}\def\b{\beta}\def\S{\Sigma}\def\C{\cite}
\def\r{\ref}\def\ba{\begin{eqnarray}}\def\ea{\end{eqnarray}}
\def\n{\nonumber}\def\R{\rho}\def\X{\Xi}\def\x{\xi}\def\la{\lambda}
\def\d{\delta}\def\s{\sigma}\def\f{\frac}\def\D{\Delta}\def\pa{\partial}
\def\Th{\Theta}\def\o{\omega}\def\O{\Omega}\def\th{\theta}\def\ga{\gamma}
\def\Ga{\Gamma}\def\t{\times}\def\h{\hat}\def\rar{\rightarrow}
\def\vp{\varphi}\def\inf{\infty}\def\le{\left}\def\ri{\right}
\def\foot{\footnote}\def\vep{\varepsilon}\def\N{\bar{n}(s)}
\def\k{\kappa}\def\sq{\sqrt{s}}\def\bx{{\mathbf x}}\def\La{\Lambda}
\def\bb{{\bf b}}\def\bq{{\bf q}}\def\cp{{\cal P}}\def\tg{\tilde{g}}
\def\cf{{\cal F}}\def\bN{{\bf N}}\def\Re{{\rm Re}}\def\Im{{\rm Im}}
\def\bk{\hat{\mathbf{k}}}\def\cl{{\cal L}}\def\cs{{\cal S}}\def\cn{{\cal N}}
\def\cg{{\cal G}}\def\q{\eta}\def\ct{{\cal T}}\def\bbs{\mathbb{S}}
\def\bU{{\mathbf U}}\def\bE{\hat{\mathbf e}}\def\bc{{\mathbf C}}
\def\vs{\varsigma}\def\cg{{\cal G}}\def\ch{{\cal H}}\def\df{\d/\d }
\def\mz{\mathbb{Z}}\def\ms{\mathbb{S}}\def\kb{\hat{\mathbb
K}}\def\cd{\mathcal D}\def\mj{\mathbf{J}}\def\Tr{{\rm Tr}}
\def\bu{{\mathbf u}}\def\by{{\mathrm y}}\def\bp{{\mathbf p}}
\def\k{\kappa} \def\cz{{\mathcal Z}}\def\ma{\mathbf{A}}
\def\me{\mathbf{E}}\def\ra{\mathrm{A}}\def\mbx{\mathbf{x}}
\def\ru{\mathrm{u}}\def\rP{\mathrm{P}}\def\rp{\mathrm{p}}\def\z{\zeta}
\def\my{\mathbf Y}\def\ve{\varepsilon}\def\bw{\mathbf
W}\def\hp{\hat{\p}}\def\hh{\hat{h}}\def\hx{\hat{\x}}\def\hk
{\hat{\kappa}}\def\hj{\hat{j}}\def\eb{\mathbf{e}}\def\bj{\mathbf{j}}
\def\T{\tau}\def\by{\bold y}\def\mH{\mathcal{H}}\def\mbk{\mathbf{k}}

\begin{center}
\vskip 3cm {\Large\bf On phase transition signal in  VHM inelastic
collisions}

\vskip 1cm J.Manjavidze\\

\vskip 0.5cm JINR, Dubna, Russia,\\ Andronikashvili Institute of
Physics, Tbilisi, Georgia
\end{center}

\begin{abstract}
The primary intent is to show that the signal of first order phase
transition may be observed at experiment if and only if the
multiplicity is sufficiently large. We discuss corresponding
phenomenology from the point of view of experiment.
\end{abstract}

{\bf I.}{\it Introduction: the aim}

The thermodynamical approach becomes more and more popular for
description of heavy ion inelastic interactions \C{ioncol}.
Generally it based on the assumption that there is equilibrium in
a final state and the Plank distribution is useable. This is a
formal basis of the approach in frame of which the thermodynamical
parameters are introduced \C{thermod} and corresponding results
are impressive.

Here we will apply the field-theoretical approach which seems more
general and, to all appearance, it allows to introduce model-free
thermodynamic description. Namely, we will try to $derive$
thermodynamics from usual $S$-matrix theory \C{hepph}. It is
evident that there is no direct connection between equilibrium
thermodynamics and $S$-matrix theory and our primary aim is to
fined the constrains in frame of which this connection exist.

We will try to describe final state by thermodynamical parameters.
The particle distribution, as it is usual in $S$-matrix, reflects
the dynamics of interacting fields. In result we construct the
$S$-matrix interpretation of thermodynamics, or shortly the
"$S$-matrix thermodynamics". This approach is the natural
development of Wigner-functions formalism \C{wign} given in
interpretation of Carruthers and Zachariasen \C{zach}. Some ideas
of Schwinger-Keldysh \C{shw} and Niemy-Semenoff \C{niemi} was used
deriving $S$-matrix equilibrium thermodynamics.

For sake of definiteness, we will consider following quantity
\C{hepph}: \be \mu(n,s)\simeq-\f{T(n,s)}{n}\ln \s_n(s) \l{a1}\ee
to illustrate our approach. It was interpreted as {\it the work
which is necessary for a particle production}. If this is so then
$\mu(n,s)$ must be sensitive to the first order phase transition.
Indeed, if the state is unstable against particle production then
$\mu(n,s)$ must decrease with number of particles. We will discuss
this idea latter.

We will consider

--- the physical meaning of $\mu(n,s)$ \\and

--- how it can be measured.\\ It must be noted that r.h.s. of
definition (\r{a1}) contains $\sim(1/n)$ corrections. Therefore
only the very high multiplicity (VHM) events will be considered.

It must be noted that in VHM region the dynamical models (Regge,
LLA QCD) can not be used \C{hepph} and the thermodynamical
description is mostly useful.

It is important that $\mu(n,s)$ is defined by the directly
measurable quantities:

--- $n$ is the multiplicity;

--- $T(n,s)$ is the mean energy of produced particles;

--- $\s_n(s)$ is the multiplicity distribution normalized on unite.\\
Despite the fact that all this quantities are simple and directly
measurable we must discuss them in more details. The point is that
above definition of $\mu$ have physical, experimental, meaning in
the frame of some restrictions.

{\bf II.} {\it Multiplicity}

By definition $n$ is the $measured$ multiplicity. This means that
it may include:

--- only the observed in given experiment charged particles;

--- particles in restricted range of rapidity. \\ This feature
makes the task of measurement of $\mu(n,s)$ real.

But there is difficulty: multiplicity must be sufficiently large.
I will explain this restriction somewhat later.

{\bf III.} {\it Temperature}

The temperature is first important parameter. In $S$-matrix
approach it is introduced as the Lagrange multiplier of energy
conservation law. From experimental point of view this means that
$T(n,s)$ is the mean energy of produced particles. But considering
event-by-event measurement the mean energy is the fluctuating
quantity. Therefore, $T$ would have physical meaning if and only
if fluctuations of $T$ are Gaussian. It must be underlined that
this is the $necessary$ condition since formally the expansion of
cross section near $T(n,s)$ have $zero$ convergence radii. In
result $T(n,s)$ have physical meaning if and only if the
inequalities \C{hepph}: \be \le. \f{<\prod_{i=1}^l(\e_i-<\e>)>}
{<\prod_{i=1}^2(\e_i-<\e>)>} \ri|_{n,s}<<1,~~l=3,4,..., \l{a5}\ee
are satisfied. Here $\e_i$ is the energy of $i$-th particle and
averaging is performed over all events at given multiplicity $n$
and energy $s$. This condition will be used for definition of VHM
region.

The ratio of correlators (\r{a1}) at $l=3$ was investigated
recently using CDF data and it was shown that it falls down with
$n$ for $T\sim 1 GeV$. This result will be published soon.

One can say that if (\r{a5}) is hold than the system is in {\it
thermal equilibrium} state. This means that particle energy
spectra is Boltzmann-like and is defined by one parameter,
$T(n,s)$. Such state is necessary to observe phase transition. But
it is not clear is this condition sufficient to use the
equilibrium thermodynamics formalism? We will see that there is
also additional condition.

{\bf IV.} {\it Cross section}

Last quantity is the multiple production cross section $\s_n(s)$.
Introducing temperature, having equilibrium condition one may
consider $\s_n$ as the $n$-particle partition function. This
assumption means that exist the set of condition in frame of which
we can set $\s_n$ equal to partition function of equilibrium
thermodynamics.

It must stressed that only the $equilibrium$ would be
considered since this case is mostly simple and in the time may be
excluded from the
field-theoretical formalism following to ergodic hypothesis. That
requirement leads to the following condition: \be
T(n,s)<<m, \l{3}\ee where the produced hadron mass, $m$, is a
natural scale parameter. It must be noted that (\r{3}) is the pure
theoretical condition since in this case one may apply the
low-temperature expansion. In its frame $\s_n$ is defined by the
partition function \be \R=Tr[e^{-\b H(\vp)+\b\int j\vp}], ~~
\b=1/T,~~H=\int d^3x h \l{4'}\ee where $h$ is the Hamiltonian
density and and $j$ is the source of external particles. Notice
that $\vp=\vp(\bx)$ is the time independent field. Low-temperature
expansion is considered in \C{lowtemp}.

Inequality (\r{3}) impose definite limitations for experiment:

--- {\bf the intermediate incident energy experiments are mostly
useful;}

--- {\bf the multiplicity must be sufficiently high;}

--- {\bf the central rapidity region, where the momentum of produced particles
is small, must be considered.}\\ We will assume that this
requirements can be hold. At the end let us calculate $\mu(n,s)$
defined in (\r{a1}).

{\bf V.} {\it Example: phase space integral}

Let us assume that the particle production dynamics is restricted
only by the energy-momentum conservation law. It must be noted
that if $n>>1$ then the ration (\r{a5}) is small in considered
model, $\sim n^{(2-l)/2}\to0,~l=3,4,...$ Therefore $T(n,s)$ have
physical meaning.

{\bf A.} In the case of non-relativistic particles production the
phase space is occupied densely and the temperature decease with
$n$: \be T(n,s)\simeq\f{2m(n_{max}-n)}{3n}<<O(1/n)
,~~~n_{max}=\sqrt{s} /m,\l{4}\ee but chemical potential \be
\mu(n,s)\simeq\f{m} {n_{max}-n}\ln\f{1}{n_{max}-n}\l{5}\ee
increase unrestrictedly when $(n_{max}-n)\to0$. This is
consequence of dense distribution of particle in the phase space.

{\bf B.} Unfortunately I can not give closed expression of the
case when momentum of particle $|k|\sim m$. In the case of
$|k|>>m$ one can find: \be T(n,s)\simeq \sqrt{s}/n\l{6}\ee and \be
\mu(n,s)\simeq\f{3}{4}m\f{n_{max}}{n}\ln\f{n_{max}}{2n},~~1<<n<<
n_{max}\l{7}\ee decrease with $n$. This is a consequence of empty
phase space.

Resulting picture given on Fig.1.

{\bf VI.} {\it Example: Ising model}

The interaction can be simply included if the lowest order of
low-temperature expansion is considered. Having (\r{4'}) one may
consider various model Hamiltonian in the frame of Ising model
\C{langer}, when \be h=\pi^2/2+(\nabla
\vp)^2/2+m(T-T_c)\vp^2/2+\la\vp^4/4,~ \la>0,\l{}\ee see for
details \C{lowtemp}. It is important that in VHM region one may
use the semiclassical approximation.

{\bf A. Stable ground state, $T>T_c$,} Fig.2. \\In that case the
temperatures decrease with multiplicity: \be T(n,s)/m\sim
n^{2/3}\l{8}\ee but chemical potential increase with $n$: \be
\mu(n,s)/m\sim n^2.\l{11}\ee Therefore, the stable vacuum leads to
repulsion and production of one additional particle needs
additional work. This is natural explanation of (\r{11}), Fig.3.

{\bf B. Unstable ground state, $T<T_c$,} Fig.4. \\In this case the
temperature tends to its critical value: \be T(n,s)\sim
T_c(1-\ga/n^4) \l{}\ee and chemical potential decrease with $n$,
Fig.5: \be \mu(n,s)\sim T_c(1-\ga/n^4)/n^5.\l{}\ee This result has
evident explanation: the boiling state is unstable against
particles evaporation.

Therefore, interaction may drastically change $n$ dependence of
thermodynamical parameters.

{\bf VII.} {\it Conclusions}

One may conclude:

--- $\mu(n.s)$ is sensitive to the interaction character: it
decrease if the ground state is unstable.

--- considering the very high multiplicity events one my hope to
investigate collective phenomena in hadron system.

\begin{figure}[c]%{14em} \vglue -0.2em \hglue -1em
\begin{center}
\includegraphics[width=20em]{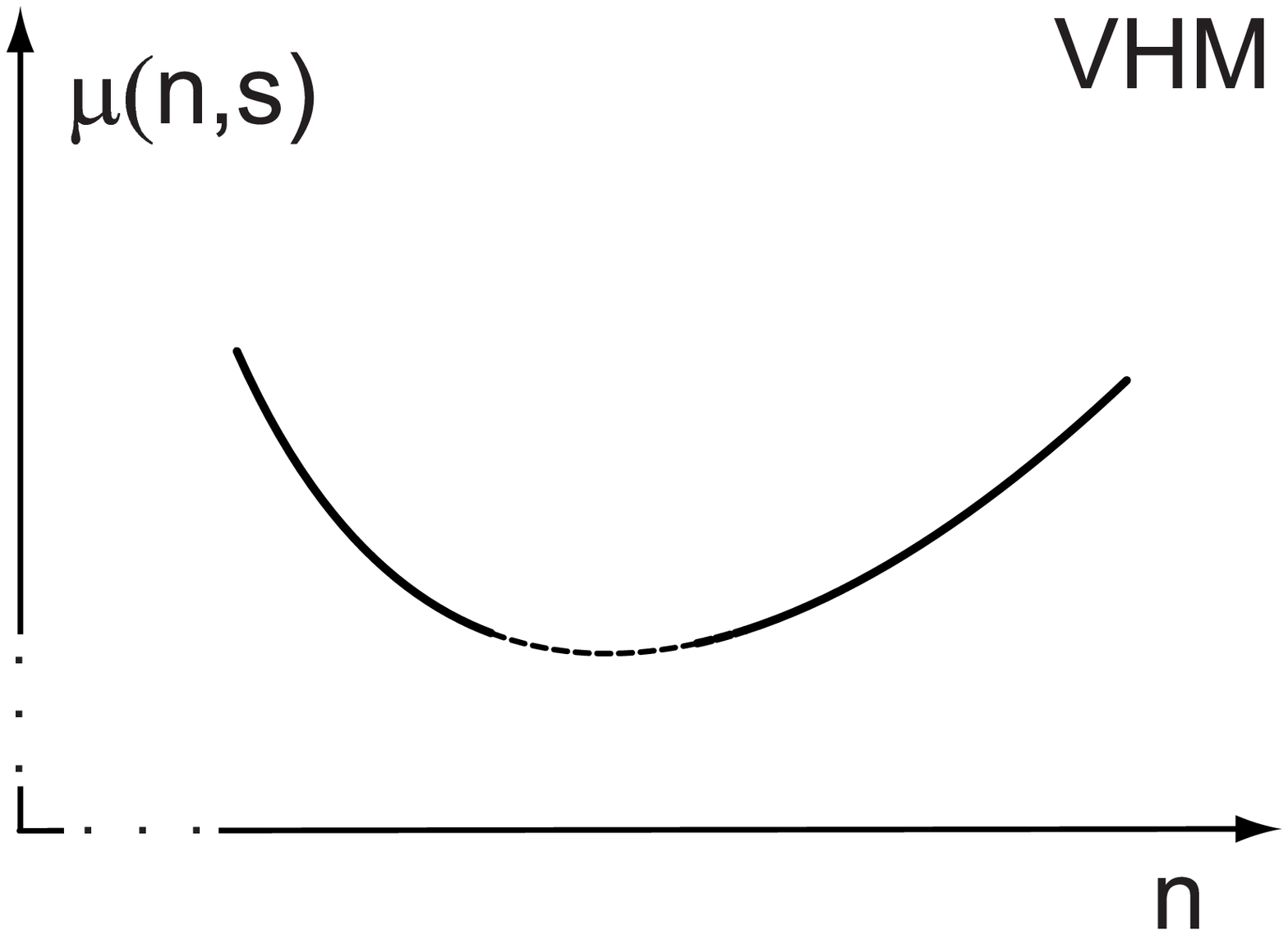}
\end{center}
\vglue 0em \caption{\footnotesize Pure phase space.} %\vglue 0em
\end{figure}

\begin{figure}[c]%{14em} \vglue -0.2em \hglue -1em
\begin{center}
\includegraphics[width=20em]{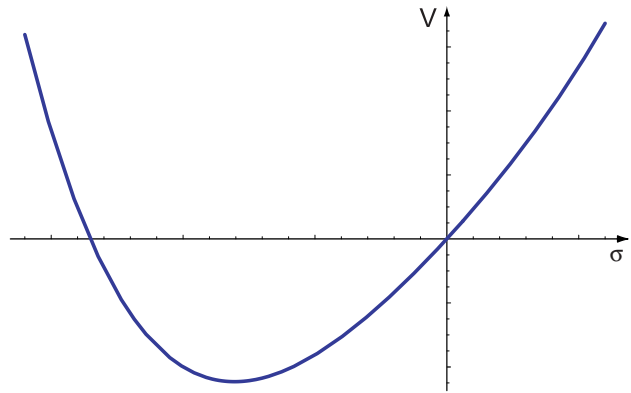}
\end{center}
\vglue 0em \caption{\footnotesize Stable ground state disturbed by
$j$.} %\vglue 0em
\end{figure}

\begin{figure}[c]%{14em} \vglue -0.2em \hglue -1em
\begin{center}
\includegraphics[width=20em]{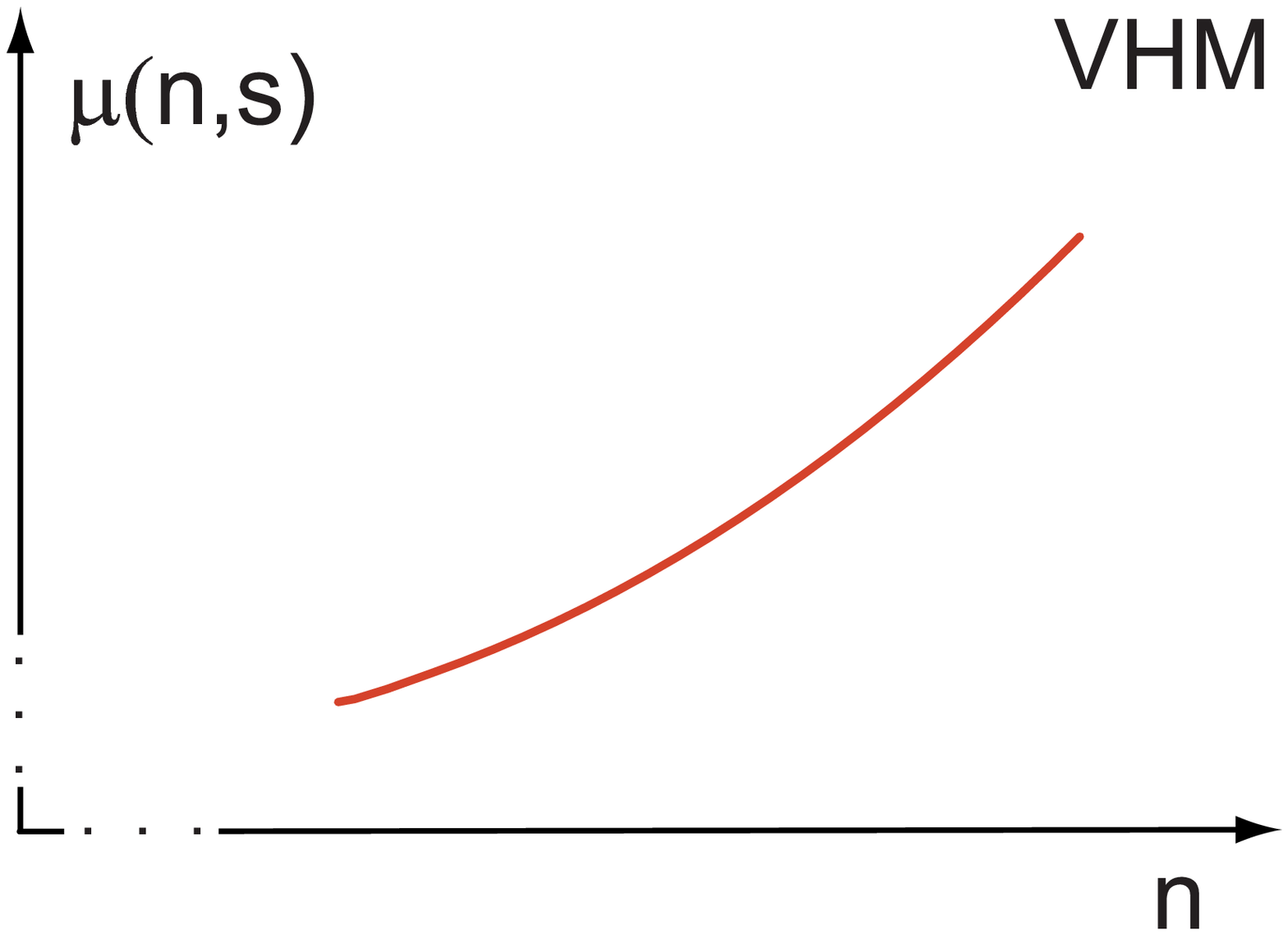}
\end{center}
\vglue 0em \caption{\footnotesize Stable ground state.} %\vglue 0em
\end{figure}

\begin{figure}[c]%{15em} \vglue -0.2em \hglue -0.5em
\begin{center}
\includegraphics[width=20em]{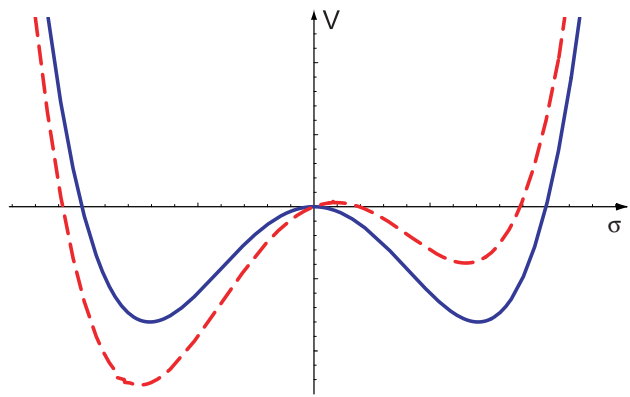}
\end{center}
%\vglue 0em
\caption{\footnotesize Solid line: undisturbed by $j$
potential and dotted line includes $j$.} %\vglue 1em
\end{figure}

\begin{figure}[c]%{14em} \vglue -0.2em \hglue -1em
\begin{center}
\includegraphics[width=20em]{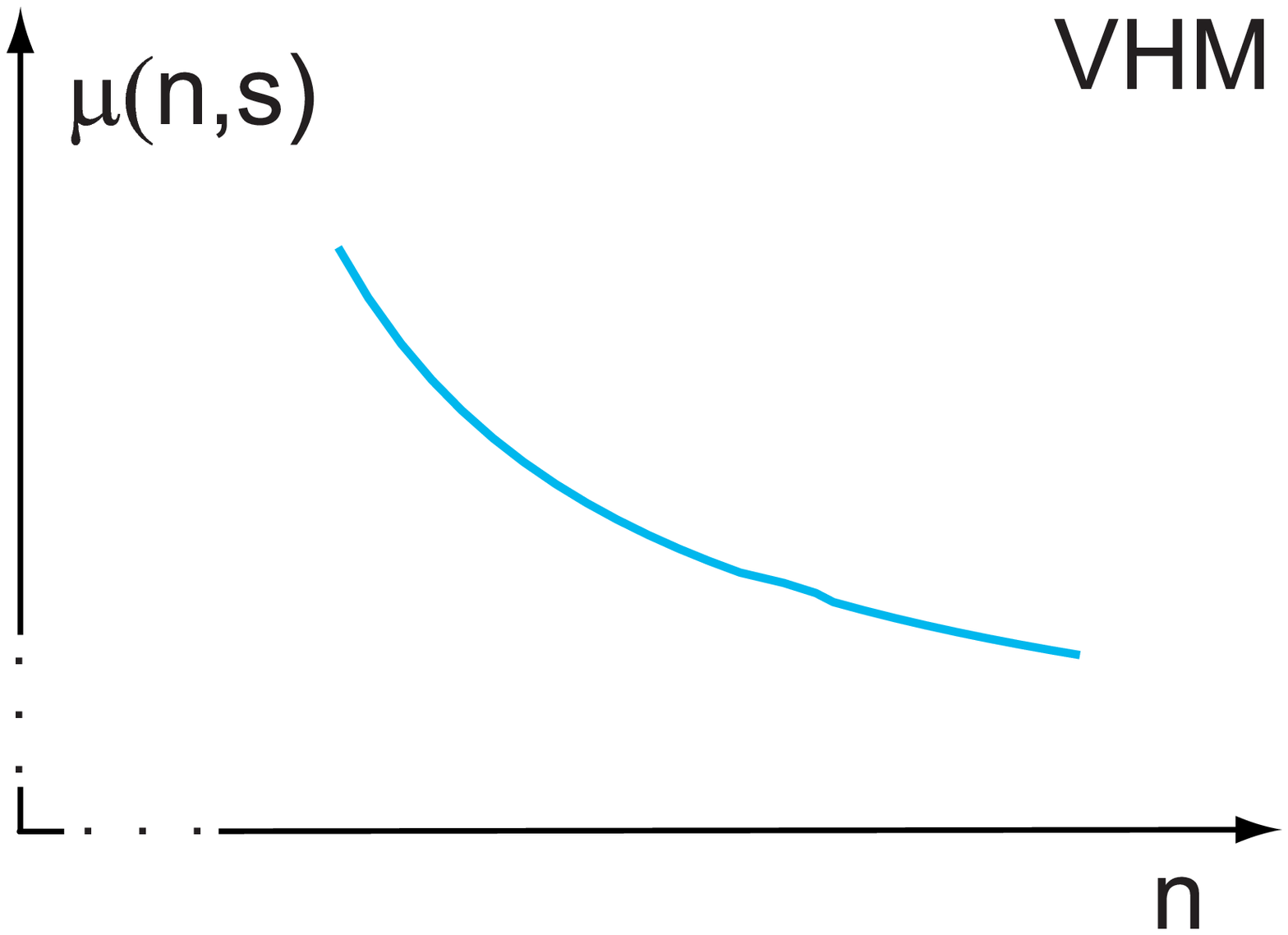}
\end{center}
\vglue 0em \caption{\footnotesize Unstable ground state.} %\vglue 0em
\end{figure}


\begin{thebibliography}{99}

\bibitem{ioncol}  BNL Report, {\it Hunting the Quark Gluon Plasma},
BNL-73847-2005; C.Alt et al., The NA-49 Collaboration,
nucl-ex/0710.0118;

\bibitem{thermod}  M.Creutz, Phys. Rev., D15(1977)1128; M.Gazdzicki and
M.I.Gorenstein, Acta Physica Polonica, B 30, (1999)2705; A.N.
Sissakian, A.S. Sorin, V.D. Toneev (Dubna, JINR) Talk given at
33rd International Conference on High Energy Physics (ICHEP 06),
Moscow, Russia, 26 Jul - 2 Aug 2006. nucl-th/0608032;
P.Braun-Munzinger, K.Redlich and J.Stachel, nucl-ph/0304013;
A.Adronic, P.Braun-Munzinger and J.Stachel, Nucl.pHys.,
A772(2006)167

\bibitem{hepph}  J.Manjavidze and A.Sissakian,, {\it Proc. VHM
Physics Workshops}, ed. A.Sissakian and J.Manjavidze (World
Scient., 2008), hep-ph/0808.2571


\bibitem{wign} E.Wigner, Phys. Rev., 40(1932)749;
K.Hisimi, Proc. Phys. Math. Soc. Jap., 23(1940)264; R.J.Glauber,
Phys. Rev. Lett., 10 (1963) 84; E.C.G.Sudarshan, Phys. Rev. Lett.,
10(1963)177; R.E.Cahill and R.G.Glauber, Phys. Rev.,
177(1969)1882; S.Mancini, V.I.Man'ko and P.Tombesi, Quant. Semicl.
Opt., 7(1995)615; V.I.Man'ko, L.Rose and P.Vitale, hep-th/9806164

\bibitem{zach} P.Carruthers and F.Zachariasen, Phys.Rev., D13 (1986)
950; P.Carruthers and F.Zachariasen, Rev.Mod.Phys., 55 (1983) 245

\bibitem{shw} J.Schwinger, J.Math.Phys., A9 (1994) 2363;
L.Keldysh, Sov.Phys. JETP, 20(1964)1018; P.M.Bakshi and
K.T.Mahanthappa, J.Math.Phys., 4(1961)1; {\it ibid.}, 4(1961)12

\bibitem{niemi} A.J.Niemi and G.Semenoff, Ann.Phys. (NY), 152 (1984)
105; N.P.Landsman and Ch.G.vanWeert, Phys.Rep., 145(1987)141

\bibitem{physrep} J.Manjavidze and A.Sissakian, Phys. Rep.,
346(2001)1, hep-ph/0105245

\bibitem{lowtemp} J.Manjavidze and A.Sissakian, to be published

\bibitem{langer} T.D.Lee and C.N.Yang, Phys.Rev., 87(1952)404,
410; J.S.Langer, Ann.Phys., 41(1967)108

\end{thebibliography}
\end{document}